\documentclass[preprint]{aastex}

\slugcomment{}

\shorttitle{NIR Coronagraphic Observations of UY Aur}
\shortauthors{Hioki et al.}

\begin{document}


\title{Near-Infrared Coronagraphic Observations \\of the T Tauri Binary System UY Aur\altaffilmark{1}}


\author{Tomonori Hioki\altaffilmark{2}, Yoichi Itoh\altaffilmark{2}, Yumiko Oasa\altaffilmark{2}, Misato Fukagawa\altaffilmark{3}, 
\\Tomoyuki Kudo\altaffilmark{4}, Satoshi Mayama\altaffilmark{4}, Hitoshi Funayama\altaffilmark{2}, Masahiko Hayashi\altaffilmark{4,5}, 
\\Saeko S. Hayashi\altaffilmark{4,5}, Tae-Soo Pyo\altaffilmark{5}, Miki Ishii\altaffilmark{5}, Takayuki Nishikawa\altaffilmark{4}, and Motohide Tamura\altaffilmark{6}}




\altaffiltext{1}{Based on data collected at the Subaru Telescope, which is operated by the National Astronomical Observatory of Japan.}
\altaffiltext{2}{Graduate School of Science, Kobe University, 1-1 Rokkodai, Nada-ku, Kobe 657-8501}
\email{otopen@kobe-u.ac.jp}
\altaffiltext{3}{Division of Particle and Astrophysical Sciences, Nagoya University, Furo-cho, Chikusa-ku, Nogoya 464-8602}
\altaffiltext{4}{School of Mathematical and Physical Science, The Graduate University for Advanced Studies, 2-21-1 Osawa, Mitaka, Tokyo 181-8588}
\altaffiltext{5}{Subaru Telescope, National Astronomical Observatory of Japan, 650 North A'ohoku Place, Hilo, HI96720, USA}
\altaffiltext{6}{Optical and Infrared Astronomy Division, National Astronomical Observatory of Japan, 2-21-1 Osawa, Mitaka, Tokyo 181-8588}


\begin{abstract}
We present a near-infrared image of UY Aur, a  $0\farcs9$ separated binary system, using the Coronagraphic Imager with Adaptive Optics on the Subaru Telescope. 
Thanks to adaptive optics, the spatial resolution of our image was $\sim 0\farcs1$ in the full width at half maximum of the point spread function, the highest achieved. 
By comparison with previous measurements, we estimated that the orbital period is $\sim 1640 \pm 90$ yrs and the total mass of the binary is $\sim 1.73 \pm 0.29$ $M_{\sun}$. 
The observed $H$-band magnitude of the secondary varies by as much as 1.3 mag within a decade, while that of the primary is rather stable. 
This inconstancy may arise from photospheric variability caused by an uneven accretion rate or from the rotation of the secondary. 
We detected a half-ring shaped circumbinary disk around the binary with a bright southwest part but a barely detectable northeast portion. 
The brightness ratio is $\gtrsim 57 \pm 5$. 
Its inner radius and inclination are about 520 AU and $42\degr \pm 3\degr$, respectively. 
The disk is not uniform but has remarkable features, including a clumpy structure along the disk, circumstellar material inside the inner cavity, and an extended armlike structure. 
The circumstellar material inside the cavity probably corresponds to a clump or material accreting from the disk onto the binary. 
The armlike structure is a part of the disk, created by the accretion from the outer region of the disk or encounters with other stellar systems. 
\end{abstract}




\section{Introduction}

Many T Tauri stars have protoplanetary disks of gas and dust where planets are born. 
The disks have been investigated in various wavelengths. 
However, the observations have focused mainly on the disk around single stars, while more than half of T Tauri stars are binaries (Ghez et al., 1993; Leinert et al., 1993). 
Artymowicz $\&$ Lubow (1996) predicted that a binary system has two kinds of disks: a circumstellar disk associated with each star and a ring-shaped circumbinary disk around the binary system. 
A cavity exists between the circumbinary disk and the central binary. 
A limited number of studies have examined the disks around binary (e.g., GG Tau) systems (Itoh et al., 2002; McCabe et al., 2002; Krist et al., 2005). 

UY Aur is a binary system of classical T Tauri stars (CTTSs; Duch\^ene et al., 1999) in the Taurus-Auriga star-forming region ($d \sim$ 140 pc; Elias, 1978). 
The separation between UY Aur A (primary) and B (secondary) is $0\farcs88$ (= 120 AU; Close et al., 1998; hereafter C98). 
The spectral types of UY Aur A and B are estimated to be M0 and M2.5, respectively (Hartigan $\&$ Kenyon, 2003). 

A circumbinary disk around the UY Aur binary was detected by millimeter, near-infrared, and polarimetric observations. 
Millimeter interferometric observations of $^{13}$CO emission (Duvert et al., 1998; hereafter D98) revealed that the disk is extended over $20\arcsec$ in diameter; its mass is calculated at $\lesssim$ 0.01 $M_{\sun}$. 
D98 assumed that the disk is rotating in Keplerian motion around the binary and estimated the total mass of the binary at $\sim$ 1.2 M$_{\odot}$. 
Near-infrared ($J$-, $H$-, and $K^{\prime}$- bands) observations with adaptive optics (C98) resolved the circumbinary disk with an inner cavity. 
Its inclination and inner radius were determined to be $\sim 42\degr$ and $\sim 500$ AU, respectively, assuming that the inner edge is circular. 
C98 also detected a clumpy structure along the circumbinary disk, which is expected to have a higher dust density than the surrounding area. 
$J$-band polarimetric observations revealed a centrosymmetric polarization pattern in the circumbinary dust disk (Potter et al., 2000). 
The amplitude of the polarization has a peak at a position angle (PA) = $135\degr$, which is consistent with the direction of its semimajor axis detected by C98. 
Potter et al. (2000) suggested that the disk is flat or flared (not spherical) with inhomogeneous dust density. 
As mentioned above, protoplanetary disks around binary systems have peculiar morphology. 

Here we present a high resolution image of the circumbinary disk around the UY Aur binary. 
We describe near-infrared coronagraphic observations in Section 2 and the data reduction procedures in Section 3. 
The morphology of the circumbinary disk, as well as astrometry and photometry of the binary system, are presented in Section 4. 
Finally, the results are discussed in Section 5. 

\section{Observations}

Near-infrared ($H$-band, 1.6 $\micron$) coronagraphic observations of UY Aur were carried out on 2002 November 23 and 2005 November 9 with the Coronagraphic Imager with Adaptive Optics (CIAO) on the Subaru Telescope (Tamura et al., 2000). 
A $2\farcs0$ mask in the instrument blocks the light of both UY Aur A and B (their apparent separation being $\sim 0\farcs9$). 
This mask has $\sim$ 2$\%$ transmission for the central star (Murakawa et al., 2003). 
A pupil Lyot stop reduces the diffracted light. 
The field of view is about $22\arcsec \times 22\arcsec$ with a pixel scale of $0\farcs02125$ $\mathrm{pixel^{-1}}$ and $0\farcs02133$ $\mathrm{pixel^{-1}}$ in 2002 and 2005, respectively. 

During the observations in 2005, the sky was clear and the seeing size was $\sim 0\farcs3$. 
While the Full Width Half Maximum (FWHM) of the point spread function (PSF) of the previous near-infrared observations was $\sim 0\farcs15$ (C98), the FWHM of our observations had a spatial resolution of $0\farcs09$, the highest achieved, because of adaptive optics (AO). 
The Strehl ratio was around 0.1.
We obtained 72 object frames. 
Integration time was 20 s $\times$ 1 coadd for each frame. 
After obtaining 36 frames, we performed dithering to remove hot and bad pixels. 
HD 282615 was observed as a PSF reference star before and after the UY Aur observations. 
We obtained 60 frames of the reference star, using a $0\farcs8$ mask. 
Integration time was 10 s $\times$ 3 coadds for each frame.
FS 125 was observed as a photometric calibrator (Hawarden et al., 2001). 
Twilight-flats and dark frames were taken at the end of the night.

During the observations in 2002, the seeing size was $\sim 1\farcs0$. 
We obtained 69 object frames, with an integration of 20 s $\times$ 1 coadd for each frame. 
We performed dithering after 34 frames were obtained. 
Nine frames of SAO 76648 were obtained as a PSF reference star. 
Dome-flats were taken at the beginning of the night, and dark frames were taken at the end.

\section{Data Reduction}

The object frames were calibrated using the Image Reduction and Analysis Facility (IRAF). 
The frames taken in 2002 were used only for astrometry and photometry because of unstable PSFs due to the poor seeing conditions. 
First, a dark frame was subtracted from the object frames. 
Then the frames were divided by the normalized dome-flat (taken in 2002) or twilight-flat (taken in 2005) to remove pixel-to-pixel variation in sensitivity. 
Hot and bad pixels were removed from the divided frames.
An average count in the sky region was subtracted. 
The count was measured along a concentric circle (with a radius of $\sim 8\arcsec$ and width of $\sim 1\arcsec$) on the UY Aur binary, except for the spider positions. 
The same reduction procedures were applied for the reference star frames. 

We subtracted the reference star frames from the object frames to detect faint structures buried in the halo of the central binary. 
The peak positions of the binary and the reference star were measured with the \texttt{imexamine} task for all frames. 
The object frames were shifted so that the positions of UY Aur A were centered on the image. 
The reference star frames were duplicated, and then each was shifted to adjust the peak positions of the PSFs between the reference star and each component of the UY Aur binary. 
Because the Subaru Telescope is an altazimuth telescope and the instrument rotator was used, the PA of the spider changed with time. 
The shifted reference star frames were rotated to adjust the PA of the spider to that of each object frame. 
The pixel values of the reference star frame were scaled to adjust the flux ratio between each component of the UY Aur binary, and the reference star frames were combined. 
The combined reference star frames were renormalized to match the intensities between the reference star frames and the object frames at the $0\farcs5 \times 0\farcs5$ region immediately outside the mask northwest of the binary, where the effect of the spider is negligible. 
The reference star frames were subtracted from each object frame. 
Finally, the object frames were combined. 
It should be noted that the radial profile of the PSF outside the mask is independent of both the peak position of the PSF in the mask and of the mask size. 
Artifacts of the mask residuals made by subtracting the PSF reference star are located within the mask of the object because the objects and the PSF reference stars are carefully centered in the mask. 
As a consequence, emission structures in the PSF-subtracted image are real structures or the residuals of the PSF halo due to imperfections of the AO compensation.

\section{Results}
\subsection{Circumbinary disk around the UY Aur binary}

\begin{figure}
\begin{center}
\epsscale{1.1}
\plottwo{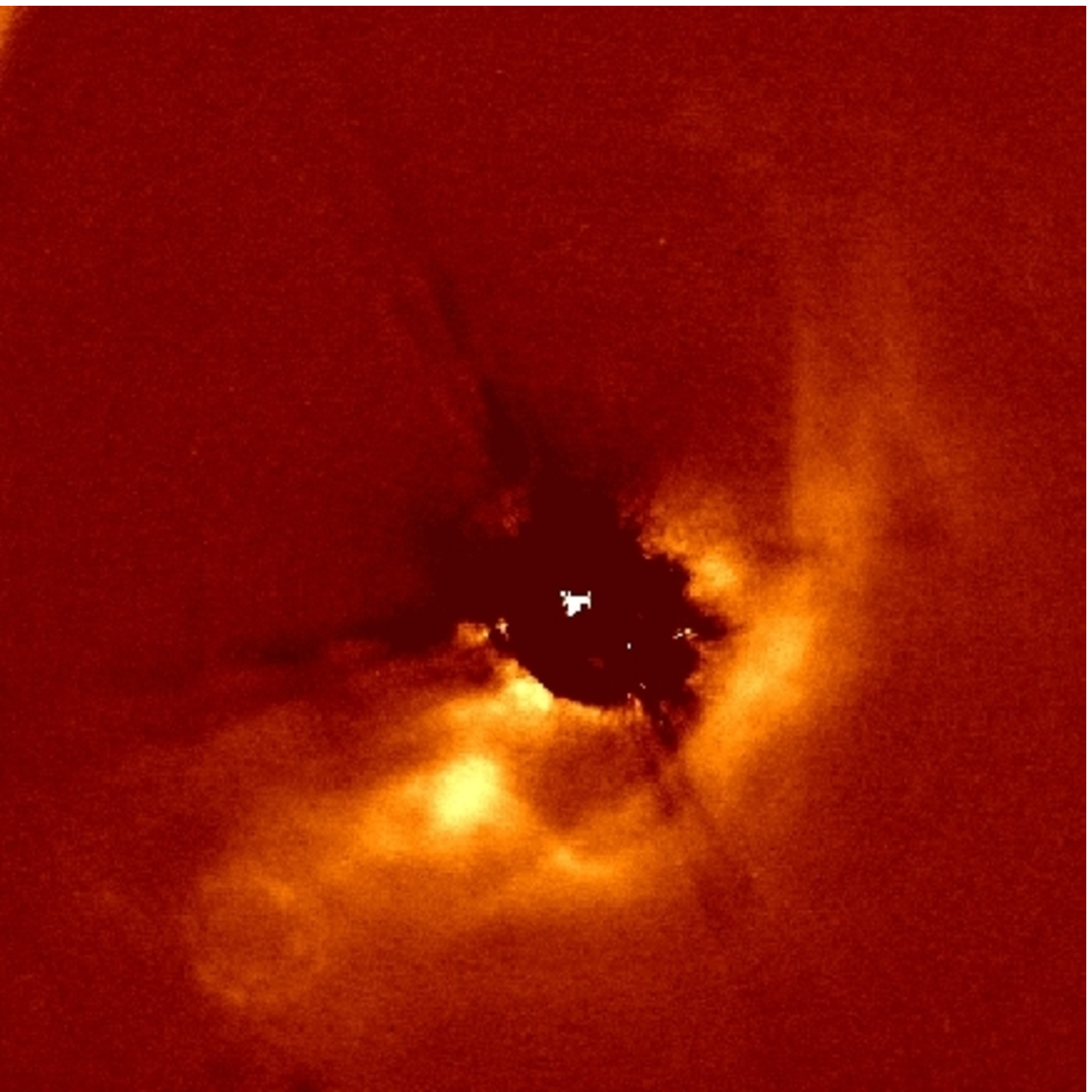}{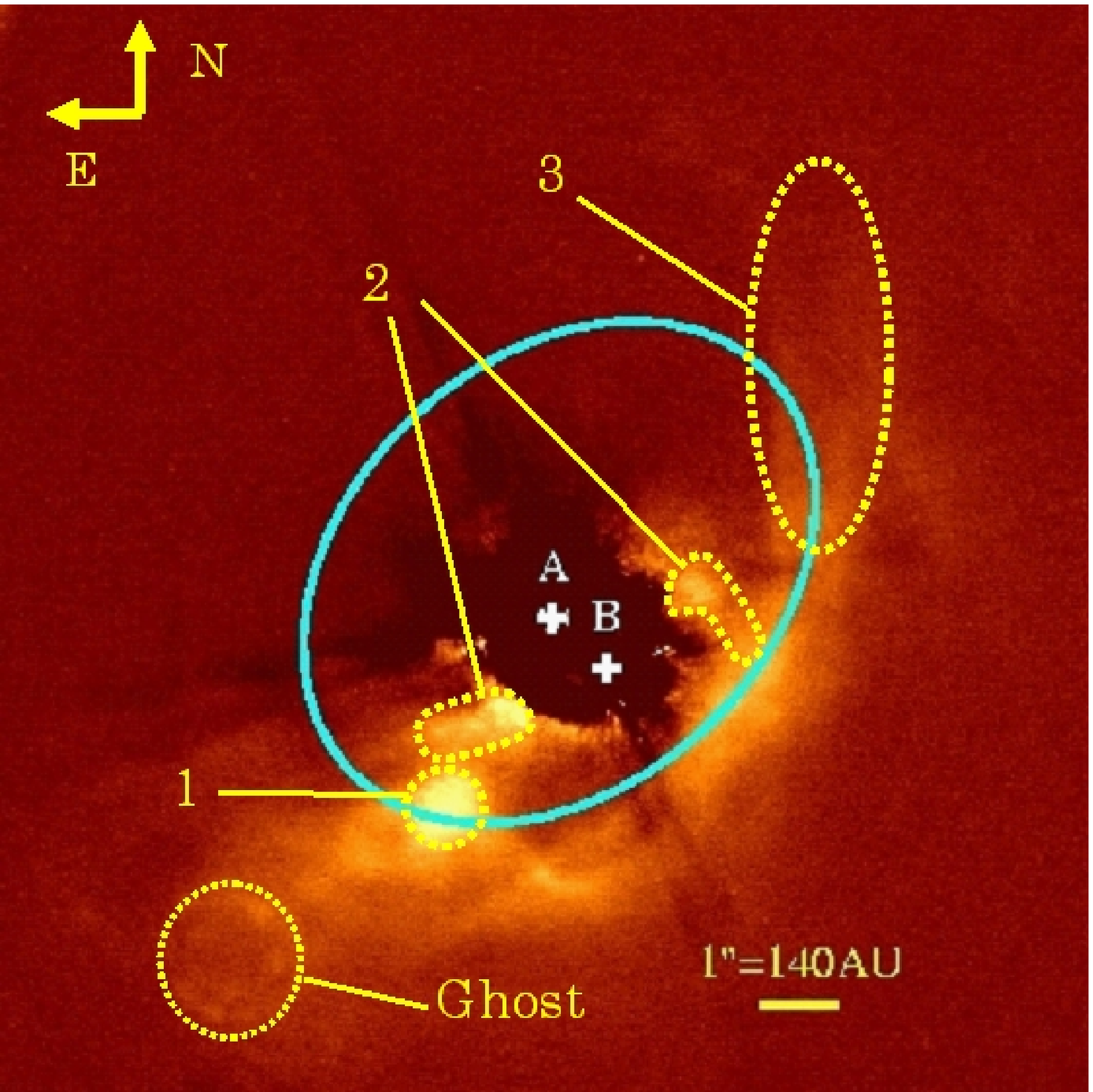}
\end{center}
\caption{$H$-band coronagraphic image of the circumbinary disk around the UY Aur binary. 
The field of view is $13\farcs2 \times 13\farcs2$. 
Diameter of the occulting mask is $2\farcs0$ (= 280 AU). 
The PSFs of the central binary are subtracted. 
The figure on the right side is the original image ($\textit{left}$) with marks. 
A and B in the mask represent the positions of the UY Aur primary and secondary, respectively. 
The southwest side of the disk is clearly seen while the northeast side is barely detected. 
The southwest side corresponds to the side nearest to us. 
The solid ellipse indicates the inner edge of the disk ($r \sim 520$ AU). 
The disk inclination $i$ was calculated to be $\sim 42\degr \pm 3\degr$. 
[1]: a clump, the brightest portion along the circumbinary disk; 
[2]: circumstellar material in the cavity; 
[3]: an extended armlike structure.}
\label{fig:fig1}
\end{figure}

Figure \ref{fig:fig1} shows the $H$-band coronagraphic image of the circumbinary disk around the UY Aur binary. 
This image was not deconvolved using the PSFs of the reference star. 
A and B represent the positions of the primary and the secondary, respectively. 
The coronagraph and AO enabled us to obtain the high spatial resolution ($\sim 0\farcs1$) image of the disk. 
If the PSFs of the reference star are subtracted too much or too little, the subtracted residual would be a circular ring. 
Because the observed disk has an elliptical shape, it is not an artifact. 
The disk has five remarkably inhomogeneous structures:
\begin{enumerate}
        \item
        A cavity between the binary and the disk. 
	\item
        Significant differences in brightness between the southwest and northeast portions of the disk. 
	\item
        A clumpy structure along the disk ([1] in Fig. \ref{fig:fig1}). 
	\item
        Circumstellar material in the cavity ([2] in Fig. \ref{fig:fig1}). 
        \item
        An extended armlike structure ([3] in Fig. \ref{fig:fig1}). 
\end{enumerate}


C98 fitted an ellipse to the peak intensity of the disk, with a semimajor axis of $\sim 500$ AU at a PA of $\sim 135\degr$ that may correspond to the inner edge of the disk. 
If the disk is circular, it is inclined by $\sim 42\degr$. 
To derive the disk parameter, we divided the southwest and southeast parts of the disk into 10 by 10 pixel regions. 
Using least-squares fitting of the peak intensity in each region, we drew an ellipse with a semimajor axis of $\sim 3\farcs7$ (= 520 AU) at a PA of $\sim 135\degr$ with an inclination of $\sim 42\degr \pm 3\degr$. 

\subsection{Astrometry}

In 2002, the apparent separation and PA between UY Aur A and B were $0\farcs886 \pm 0\farcs001$ and $229\fdg13 \pm 0\fdg04$, respectively. 
In 2005, they were $0\farcs890 \pm 0\farcs001$ and $231\fdg36 \pm 0\fdg04$, respectively. 
The binary system has an inclination of $\sim 42\degr \pm 3\degr$ relative to the line of sight, if it is coplanar with the circumbinary disk. 
As the major axis of the disk is along a PA of $\sim 135\degr$, we ascertained that the deprojected separation and PA in 2002 were $1\farcs191 \pm 0\farcs02$ and $228\fdg08 \pm 0\fdg05$, respectively. 
In 2005, they were $1\farcs194 \pm 0\farcs02$ and $229\fdg74 \pm 0\fdg05$. 

\begin{figure}
\begin{center}
\epsscale{1.7}
\plottwo{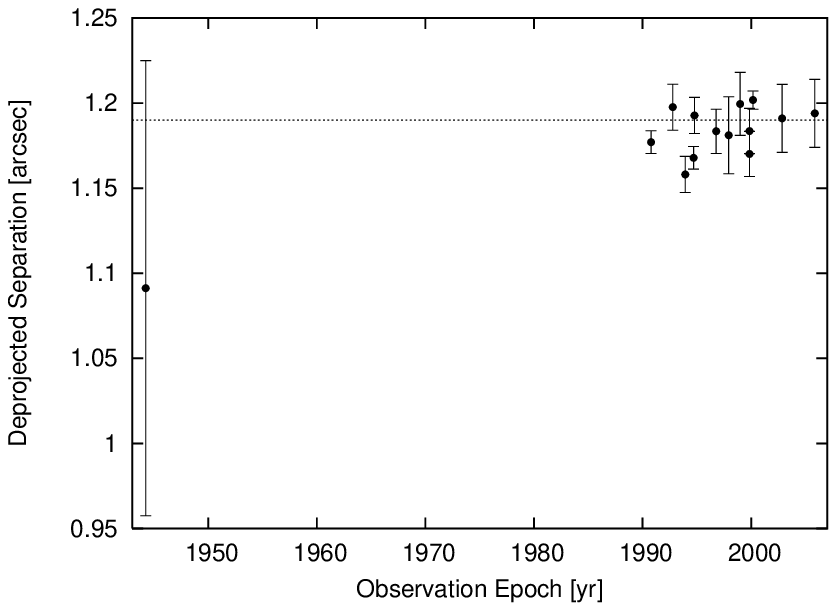}{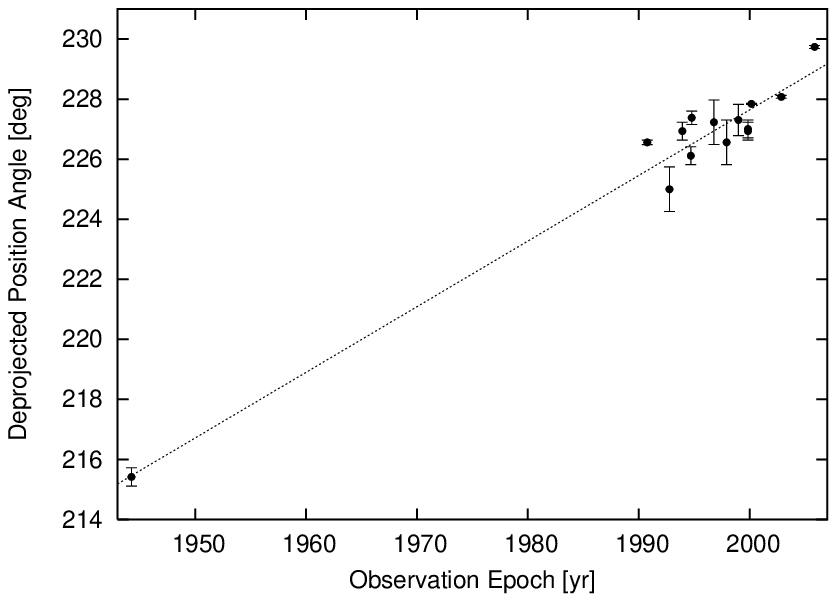}
\end{center}
\caption{$\textit{Top}$: Deprojected separation between UY Aur A and B. 
If a circular orbit is assumed, its radius is $\sim 1\farcs19$ (= 167 AU) (described by the dotted line). 
Data are from Joy $\&$ Van Biesbroeck, 1944; Ghez et al., 1993; Leinert et al., 1993; Ghez et al., 1995; C98; White $\&$ Ghez, 2001; Hartigan $\&$ Kenyon, 2003; McCabe et al., 2006; Brandeker et al., 2003; and this work (last two points). 
$\textit{Bottom}$: Deprojected position angle of the binary. 
Data are from the same sources as in Figure \ref{fig:fig2} ($\textit{top}$). 
The dotted line drawn by least-square fit indicates that UY Aur B has an angular velocity of $\sim 0\fdg22$ yr$^{-1}$ with respect to A.}
\label{fig:fig2}
\end{figure}

Figure \ref{fig:fig2} shows the deprojected separation ($\textit{top}$) and PA ($\textit{bottom}$) of UY Aur A and B. 
Comparison with previous observations shows that the separation has been constant at $1\farcs19$ (= 167 AU) since 1944. 
UY Aur B has a constant angular velocity of $0\fdg22 \pm 0\fdg01$ yr$^{-1}$ with respect to A. 
Assuming a circular orbit, the orbital period and total mass, $M\mathrm{_T}$, of the binary are calculated to be $1640 \pm 90$ yr (= $360\degr$/$0\fdg22 \pm 0\fdg01$ yr$^{-1}$) and $1.73 \pm 0.29$ $M_{\sun}$, respectively. 
The calculated $M_\mathrm{T}$ is nearly consistent with C98's result estimated from the binary motion ($1.61_{\! -0.67}^{\! +0.47}$ $M_{\sun}$), but it differs from D98's result estimated from the velocity structure of the circumbinary disk ($\sim1.2$ $M_{\sun}$). 
The assumption of a circular orbit may cause the mass difference.

\subsection{Photometry}
\label{sec:phot}

Magnitudes of UY Aur A and B were measured by aperture photometry using the \texttt{phot} task. 
Photometry was performed on the images after the reduction processes, including dark-subtraction, flat-fielding by either dome flat or twilight flat, hot and bad pixel rejection, and sky-subtraction. 
We used $0\farcs22$ and $0\farcs18$ radius apertures for the 2002 and 2005 images, respectively. 
The average intensity of the region without the circumstellar structures was set as a sky count. 
The binary magnitudes were derived from the measured magnitude under the mask and the Two Micron All-Sky Survey (2MASS) magnitude of the PSF reference star. 
Note that the mask has 2$\%$ transmission. 
In 2002, the magnitudes of UY Aur A and B were determined to be 8.11 $\pm$ 0.21 mag and 8.54 $\pm$ 0.21 mag, respectively. 
In 2005, those figures were 7.85 $\pm$ 0.10 mag and 9.55 $\pm$ 0.13 mag, respectively. 
In 1996, the values were 8.26 $\pm$ 0.07 mag and 9.85 $\pm$ 0.06 mag, respectively (C98). 
The $H$-band magnitude of UY Aur B varies up to 1.3 mag over a decade, while A is rather stable. 

\section{Discussion}
\subsection{Asymmetric brightness of the circumbinary disk}

\begin{figure}
\epsscale{1.0}
\plotone{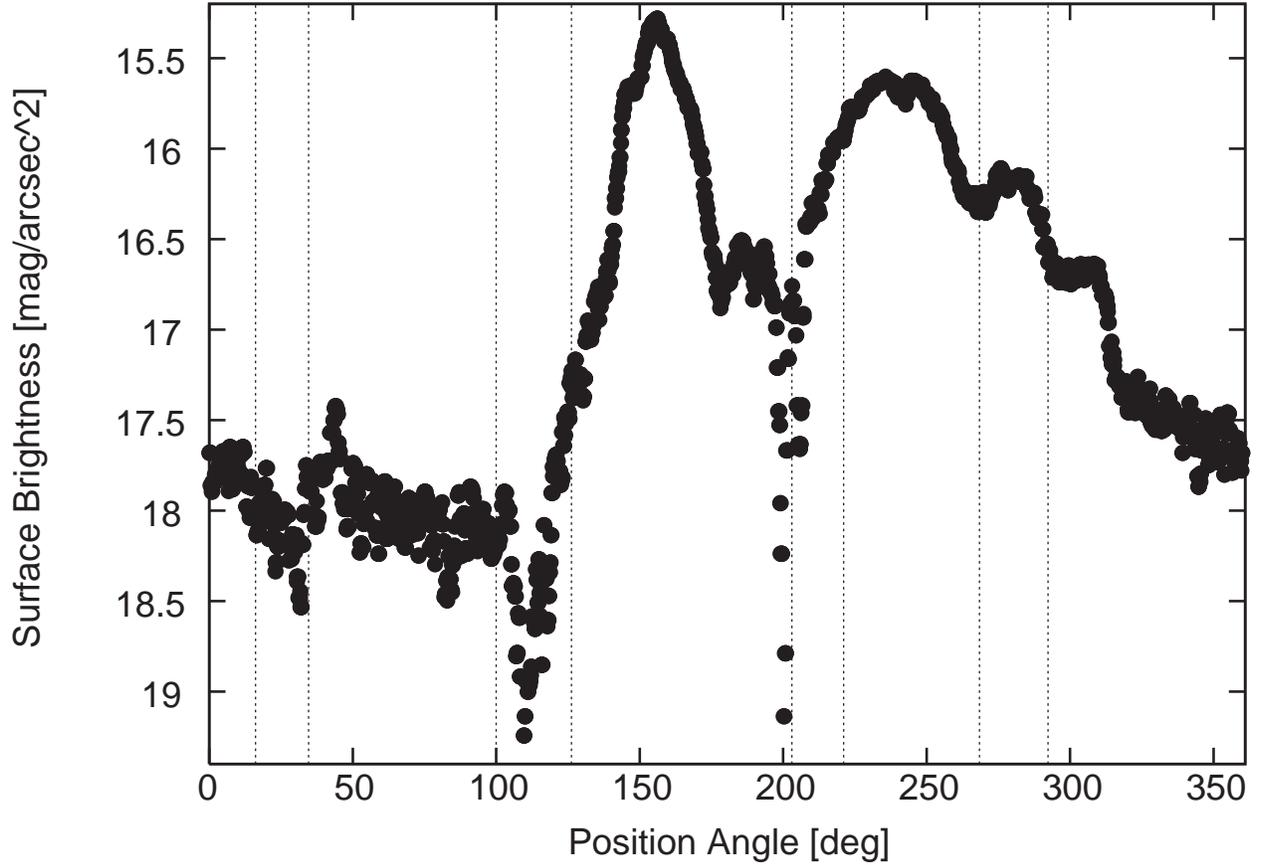}
\caption{Azimuthal intensity profile of the inner edge of the circumbinary disk. 
The points represent the average intensity of the ellipse (Fig. 1) in 3 $\times$ 3 pixel aperture. 
The intensity is normalized by the peak intensity (PA $\sim 156\degr$) in the disk, which corresponds to the clump. 
Forward scattering of the dust creates the peak in the near side ($225\degr \lesssim$ PA $\lesssim 250\degr$). 
In contrast, the far side ($45\degr \lesssim$ PA $\lesssim 70\degr$) is barely detected. 
The intensity ratio (near/far) is calculated to be $\gtrsim 57 \pm 5$. 
The vertical dotted lines represent the position angles of the spider. 
North is PA = $0\degr$, and east is PA = $90\degr$.}
\label{fig:PAvsIn}
\end{figure}

Figure \ref{fig:PAvsIn} shows the azimuthal intensity profile along the ellipse in Figure \ref{fig:fig1}, which is normalized by the peak intensity in the circumbinary disk. 
The southwest side of the disk ($225\degr \lesssim $ PA $\lesssim 250\degr$) is bright, while emission is barely detected toward the northeast portion. 
The intensity ratio of the southwest side to northeast side is $\gtrsim 57 \pm 5$, while C98 reported that the brightness ratio was $\sim 2.7$. 

We suggest that anisotropic scattering by dust causes the brightness asymmetry of the disk and that the southwest side of the disk is brighter due to the forward scattering of the dust. 
This indicates that the southwest side of the disk corresponds to the near side along our line of sight. 

\begin{figure}
\begin{center}
\epsscale{0.9}
\plotone{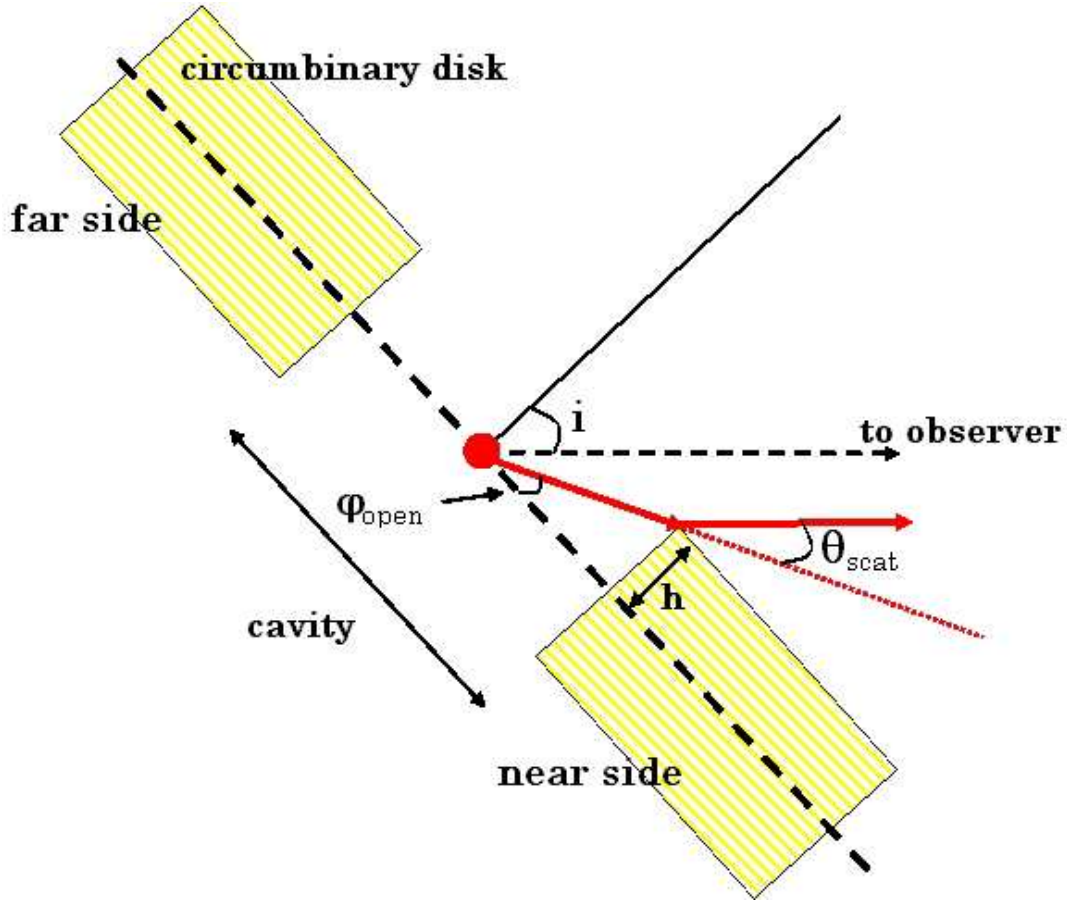}
\end{center}
\caption{Schematic view of the model of the UY Aur binary and its circumbinary disk. 
The scattering angle $\theta_{\mathrm{scat}}$, disk height $h$, and $\phi_{\mathrm{open}}$ are shown.}
\label{fig:disk}
\end{figure}

Based on this hypothesis, we described the disk brightness with a modified function of the Henyey-Greenstein scattering phase function (McCabe et al., 2002),
\begin{equation}
\Phi(\theta_\mathrm{scat}) \propto (1-g^2)(1+g^2-2g\cos{\theta_\mathrm{scat}})^{-\case{3}{2}},
\end{equation}
where $g$ is the asymmetry parameter and $\Phi(\theta_\mathrm{scat})$ is the normalized azimuthal intensity of the disk along the inner edge as a function of the scattering angle $\theta_\mathrm{scat}$. 
The scattering angle $\theta_\mathrm{scat}$ is defined as in McCabe et al. (2002),
\begin{equation}
\cos(\theta_{\mathrm{scat}} + \phi_{\mathrm{open}}) = \sqrt{1- \,\frac{1}{1+\mathrm{cos}^2(\mathrm{PA}-\mathrm{PA_0})\mathrm{tan}^2i}}(-1)^{\! j},
\end{equation}
where $\phi_{\mathrm{open}}$ is the angle related to the disk height, $h$; PA$_0$ is the position angle of the brightest part ($= 235\fdg7$); $i \sim 42\degr$ is the disk inclination (see Figure \ref{fig:disk}). 
$j = 1$ for cos($\mathrm{PA}-\mathrm{PA_0}$) $>$ 0, otherwise $j = 0$. 
The scattering angle $\theta_\mathrm{scat}$ has minimum at PA = PA$_0$. 

\begin{figure}
\begin{center}
\epsscale{1.7}
\plottwo{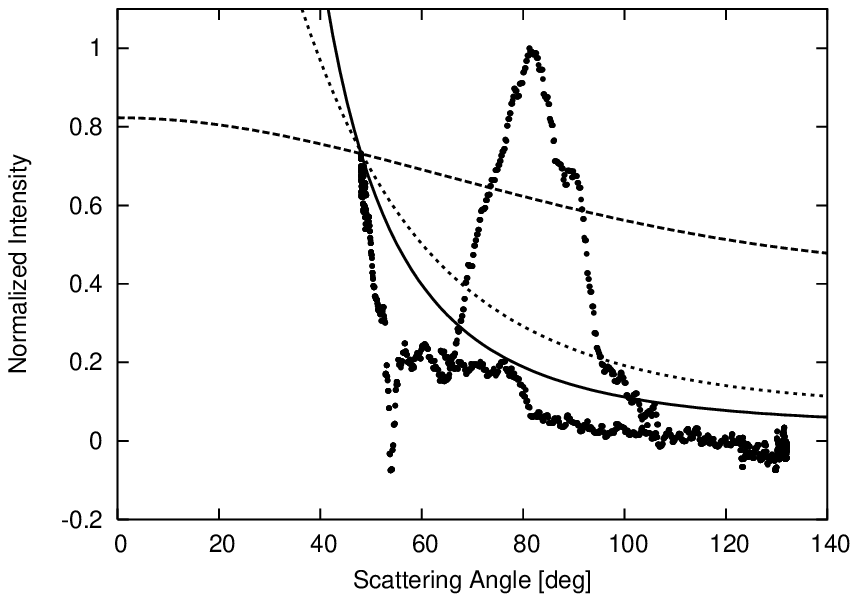}{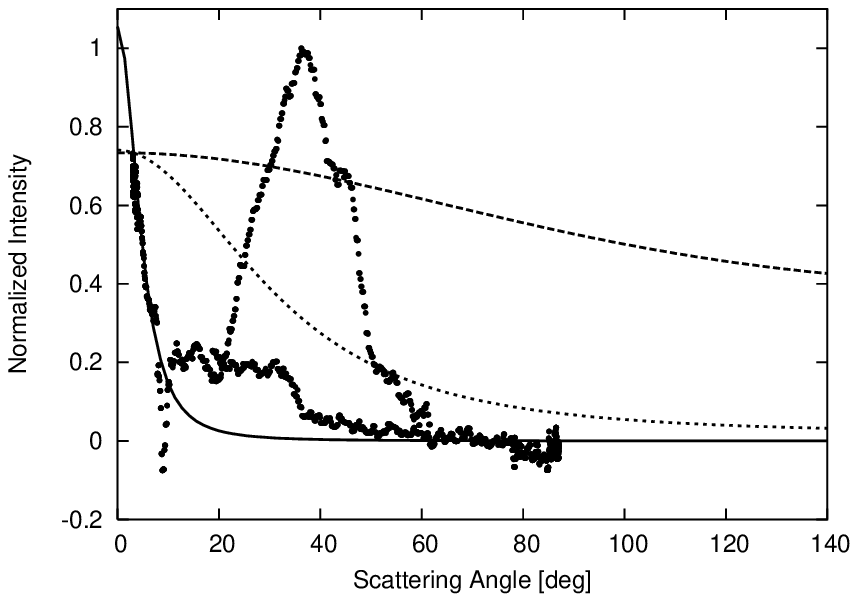}
\end{center}
\caption{The observed azimuthal intensity profile of the circumbinary disk ($\textit{points}$) as a function of scattering angle $\theta_{\mathrm{scat}}$. 
Lines represent the Henyey-Greenstein function (dashed lines for $g = 0.1$, dotted lines for $g = 0.5$, and solid lines for $g = 0.9$). 
The small scattering angle corresponds to the near side of the disk while the large angle corresponds to the far side. 
The local peaks at $\theta_{\mathrm{scat}}$ = $65\degr$--$95\degr$ ($\textit{top}$) and $20\degr$--$50\degr$ ($\textit{bottom}$) represent the clump. 
$\textit{Top:}$ Flat disk ($h = 0$). 
The function does not fit well to the observed profile with any $g$. 
$\textit{Bottom:}$ Geometrically thick disk ($h = 520$ AU). 
The observed profile is fitted by the Henyey-Greenstein function with $g = 0.9$.}
\label{fig:g}
\end{figure}

The normalized intensity $\Phi(\theta_\mathrm{scat})$ is presented in Figure \ref{fig:g}. 
In the case of a geometrically thin disk (i.e., $\phi_{\mathrm{open}} = 0$; Fig. \ref{fig:g}, $\textit{top}$), the observed intensity profile is not well reproduced with any $g$. 
In the case of $\phi_{\mathrm{open}} = 45\degr$ (i.e., $h = 520$ AU; Fig. \ref{fig:g}, $\textit{bottom}$), the observed profile is fitted with $g \sim 0.9$. 
However, it is hard to imagine a circumbinary disk with such a large asymmetry parameter of the particles and such huge disk height. 
The brightest portion in the circumbinary disk may be a local peak of dust distribution. 
Otherwise, we propose a situation in which the near side is free from heavy extinction while the other area is deeply embedded. 
We cannot exclude the possibility that the extinction varies across the disk and induces a variation in the brightness observed across the disk. 

\subsection{Mass of clump}

\begin{figure}
\begin{center}
\epsscale{0.7}
\plotone{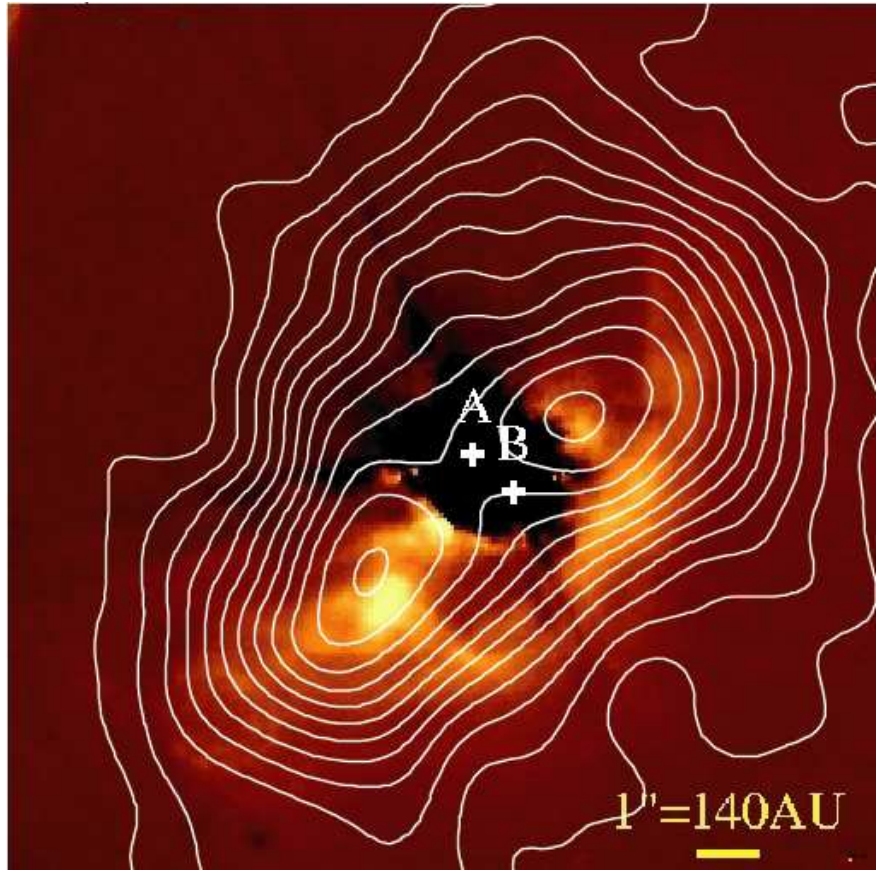}
\end{center}
\caption{$H$-band image overlaid $^{13}$CO $(J$ = 2 $\to$ 1) integrated intensity map (D98; $\textit{contour}$). 
The spatial resolution in $^{13}$CO ($J$ = 2 $\to$ 1) is $2\farcs9 \times 2\farcs3$. 
The field of view is $14\farcs4 \times 14\farcs4$. 
The gas distribution has a peak near the near-infrared clump. 
$^{13}$CO emission extends over $20\arcsec$ (= 2800 AU) in diameter.}
\label{fig:CO}
\end{figure}

The clump is detected along the disk southeast of the binary (PA $\sim156\degr$) and seems to be circular. 
Its diameter is $\sim 1\farcs3$ (= 180 AU) in the FWHM, which is significantly larger than the FWHM of the PSF ($\sim 14$ AU). 
The apparent distance between the clump and the center of the gravity of the binary is $\sim 2\farcs9$ (= 410 AU). 
We calculated its surface brightness to be 15.3 $\pm$ 0.1 mag arcsec$^{-2}$, using an aperture diameter of $\sim 1\farcs3$ and the same sky count as the binary photometry ($\S$ \ref{sec:phot}). 
The clump was also resolved by C98, who suggested that it is a local peak of the dust density with a calculated diameter of $\sim 0\farcs8$ and a surface brightness of $16.1 \pm 0.1$ mag arcsec$^{-2}$ at the $H$-band. 
By comparing our image with the $^{13}$CO-integrated intensity map (Fig. \ref{fig:CO}), we found that distribution of gas also had a peak near the near-infrared clump. 

The lower limit of the clump mass was estimated from its luminosity, assuming optically thin conditions in near-infrared. 
Following Lazareff et al. (1990), the clump luminosity $L_{\mathrm{clump}}$ is defined as: 
\begin{equation}
L_{\mathrm{clump}} = \frac{\pi r^2}{4 \pi D^2} L_{\mathrm{star}} \int e^{\! -\tau_{1}} \rho \kappa_{\! \mathrm{s}} e^{\! -\tau_{2}} du,
\end{equation}
where $L_{\mathrm{star}}$ is the luminosity of the binary, $r$ is the clump radius, $D$ is the deprojected distance between the binary and the clump ($\sim 520$ AU), $\tau_{1}$ is the optical depth between the binary and the clump, $\tau_{2}$ is the depth between the clump and the observer, 
$\rho$ is the lower limit of the clump density, $\kappa_{\! \mathrm{s}}$ is the scattering opacity of gas and dust in the clump, and $\textit{u}$ is along a light ray emitted from the binary. 
We simplify Eq. (3) with three assumptions. 
First, the light paths both from the binary to the clump and from the clump to the observer are optically thin (i.e. $e^{\! -\tau_{1}}$ and $e^{\! -\tau_{2}}$ are $\sim$ 1). 
Second, the clump is spherical, with $r\sim 90$ AU. 
Assuming uniform distribution of the dust in the clump and single scattering, the average path length of light in the clump is nearly equal to its diameter, $2r \sim 180$ AU. 
Finally, because the distance between the binary and the observer is quite large ($\sim$ 140 pc), luminosity and flux are related as 
\begin{equation}
\frac{L_{\mathrm{clump}}}{L_{\mathrm{star}}} = \frac{F_{\mathrm{clump}}}{F_{\mathrm{star}}}.
\end{equation}
With these assumptions Eq. (3) is transformed into

\begin{equation}
\rho = \frac{F_{\mathrm{clump}}}{F_{\mathrm{star}}} \frac{2D^2}{r^3 \kappa_{\! \mathrm{s}}}.
\end{equation}

We derived $F_{\mathrm{clump}} = 5.9 \times 10^{2}$ counts and $F_{\mathrm{star}} = 4.9 \times 10^{6}$ counts from the photometry above.
Given that $\kappa_{\! \mathrm{s}}$ = 18.5 cm$^{2}$ g$^{-1}$ for the mixture of graphite + silicate grains and gas (Wolfire $\&$ Cassinelli, 1986), $\rho$ is calculated to be $\sim 5.4 \times 10^{-18}$ g cm$^{-3}$ and the lower limit of the clump mass is $\sim9.6$ $M_{\earth}$. 
High spatial resolution (sub-)mm wavelength observations are required to obtain an accurate mass for the clump. 


\subsection{Circumstellar material}

Possible emissions were detected inside the cavity in the northwest and southeast areas of the UY Aur binary (hereafter called NW and SE, respectively). 
They may be spiral arms extending from the circumbinary disk to the central binary. 
C98 reported the presence of an arm at $0\farcs3$ from UY Aur B (see the $J$-band image in their Fig. 1). 
NW nearly corresponds to the position of C98's arm. 
Artymowicz $\&$ Lubow (1996) demonstrated that the secondary star perturbing a circumbinary disk produces such spiral arms. 
Otherwise, SE could be a clump. 
We calculated its lower mass limit to be $\sim 6 \times 10^{-6}$ $M_{\earth}$. 
Data reductions produced negative artifacts as strong as positive ones in the region just outside the mask ($-30\degr \lesssim$ PA $\lesssim 90\degr$), so discussions about NW and SE are less reliable. 

\subsection{Armlike structure}

The armlike structure was detected at PA $\sim 270\degr$ (see [3] in Figure \ref{fig:fig1}), which is elongated to the north and extends to $\sim 6\farcs5$ (= 910 AU) from the UY Aur binary. 
Given that the structure is a part of the circumbinary disk, we considered two possible mechanisms for its origin. 
First, cold dust may have accreted from the outer region of the circumbinary disk to the central binary. 
The outer radius of the disk was ascertained to be $\gtrsim 10\arcsec$ (= 1400 AU, D98), with the circumbinary disk having higher gas surface density than the other area (see Fig. 2 in Ochi et al., 2005). 
Second, the structure may have formed by an encounter with another stellar system. 
Pfalzner (2003) demonstrated that a close flyby of a star creates an arm structure in a circumstellar disk. 
We investigated the Taurus-Auriga star-forming region. 
Our calculation, which considered the proper motion of UY Aur and other young stellar objects (Jones \& Herbig, 1979), showed that the paths of RW Aur and JH 218 intersected with that of UY Aur during the past $10^6$ yrs. 
However, none of these stars show evidence of a stellar encounter, such as an armed disk structure. 

\subsection{Light variation of UY Aur B}
Brandeker et al. (2003) reported that the $H$-band flux ratio of UY Aur A to B changes from 4.33 $\pm$ 0.36 (1996 Oct; C98) to 1.47 $\pm$ 0.01 (2000 Feb; Brandeker et al., 2003). 
UY Aur B had a magnitude variation up to 1.3 mag at the $H$-band during the decade, while the magnitude of A was rather stable. 
This variation probably originates from the photospheric variability of UY Aur B. 
The light variation often observed in CTTSs may be caused by variable accretion rates to the central star or by rotation of a star with a hot spot on its surface. 
Simultaneous multiband observations will clarify the source of this variation. 


\section{Conclusions}
Using the Subaru/CIAO, we obtained the highest spatial resolution ($\sim 0\farcs1$) near-infrared coronagraphic observations yet achieved of the UY Aur binary. 
The southwest side of the circumbinary disk around the binary was clearly seen, while the northeast side was barely detected. 
Its inner radius and inclination are $\sim$ 520 AU and $\sim 42\degr$, respectively. 
The following disk features were observed:

\begin{enumerate}
\item
The intensity ratio (near/far) of the inner disk edge is calculated to be $\gtrsim 57 \pm 5$. 
We determined that the southwest bright side of the disk is located near to us, assuming forward scattering of the dust. 
By fitting the Henyey-Greenstein phase function given a disk height of $\sim 520$ AU, we derived the asymmetry parameter $g \sim 0.9$. 
Inhomogeneous distributions of dust in the disk and in front of the disk may also account for the huge ratio. 

\item
A clumpy structure was resolved in the southeast region of the disk. 
The apparent distance from the binary was $\sim 410$ AU. 
This is probably a dense clump of dust and gas. 
The lower limit of the clump mass is $\sim 9.6$ $M_{\earth}$, assuming optically thin conditions. 

\item
Circumstellar materials were detected at two regions within the inner cavity. 
The material in one region may have accreted from the circumbinary disk to the central binary through the cavity. 
The second region may also be a clump. 
However, these structures are located just outside the coronagraphic mask and further observations are needed. 

\item
An armlike structure northwest of the binary was probably created by dust accretion from the outer region of the disk or through a stellar encounter. 
We detected two candidates with proper motions that would have allowed them to have encountered the binary. 

\end{enumerate}

We also conducted astrometry and photometry of the binary. 
The deprojected separation between UY Aur A and B was constant at $\sim 1\farcs19$ (= 167 AU) over about 60 yrs, suggesting that the binary has a circular orbit. 
The total mass and the orbital period of the binary were $\sim 1.73 \pm 0.29$ $M_{\sun}$ and $\sim$ 1640 yrs, respectively. 
The $H$-band magnitude variation of UY Aur B was up to 1.3 mag over a decade. 
This variation may be caused by UY Aur B's photospheric variability. 

\subsection*{}
We are grateful for constructive and useful comments from an anonymous referee. 
We thank Ingrid Mann for a discussion.
This work was supported by "The 21st Century COE Program: The Origin and Evolution of Planetary Systems" of the Ministry of Education, Culture, Sports, Science and Technology (MEXT). 
Y.I. is supported by Grants-in-Aid for Scientific Research No. 16740256 of the MEXT. 
This research was partly supported by a Grant-in-Aid for Scientific Research on Priority Areas from the Ministry of Education, Culture, Sports, Science and Technology of Japan.

\newpage

\newpage

%
%
%
%

\newpage

\newpage

\newpage

\end{document}